\documentclass[pra,11pt]{revtex4-1}
\usepackage{amsmath}    
\usepackage{graphicx}   
\usepackage{verbatim}   
\usepackage{color}      
\usepackage{epsfig}
\usepackage{subfigure}  
\usepackage{hyperref}   
\newcommand{\q}{\quad}

\newcommand{\bea}{\begin{eqnarray}}
\newcommand{\eea}{\end{eqnarray}}
\newcommand{\ket}[1]{\left|{#1}\right\rangle}

\newcommand{\aver}[1]{\left\langle{#1}\right\rangle}

\newcommand{\modu}[1]{\left|{#1}\right|}

\begin{document}
\title{Signatures of entanglement in an optical tomogram}

\author{M. Rohith and C. Sudheesh}
\affiliation{Department of Physics, Indian Institute of Space Science and Technology, Thiruvananthapuram, 695 547, India.}

\begin{abstract}
We theoretically study the optical tomography of maximally entangled states generated at the output modes of a beam splitter. We consider even and odd coherent states in one of the input modes and  vacuum state in the other input mode of the beam splitter. We have shown that the signatures of entanglement can be observed directly in  the optical tomogram of the state, without reconstructing the density matrix of the system.  Two distinct types of optical tomograms are observed  in any one  of the output modes of the beam splitter  based  on the quadrature measurement in the other output mode if the output modes are entangled.  The different features shown by the optical tomograms  are verified by investigating the photon number statistics of the corresponding state.
\end{abstract}

\maketitle

\section{Introduction}
\label{introduction}
Quantum entanglement is a key resource for quantum information and quantum computing. After the EPR paper \cite{epr}, a tremendous amount of work has been done in the field of quantum entanglement \cite{Horodecki}. In most of the quantum information processes, such as, quantum teleportation \cite{bennet}, quantum cryptography \cite{nicolas}, quantum metrology \cite{vitto}, etc., the systems are prepared initially in an entangled state. Much attention is devoted to the discussion of entanglement properties of continuous variable systems, for their great practical relevance in applications to quantum optics and quantum information. Generating entangled states is one of the major tasks in quantum information processing. Various devices have been proposed and realised experimentally to generate quantum entanglement. The simplest one is by using a quantum mechanical beam splitter.

Once the entangled states are created in an experiment, it is important to characterise the state of the system precisely. Optical homodyne tomography can serve as an efficient technique to measure and reconstruct the state of entangled optical fields. Optical tomography is based on the one-to-one correspondence between the quasiprobability distribution and probability distribution of the rotated quadrature phases of the optical field \cite{vogel}. The first experimental observation of squeezed state of light, by measuring the quadrature amplitude distribution using balanced homodyne detection, has been done in \cite{smithey}. A series of homodyne measurements of the rotated quadrature operator of the field develops a quadrature histogram called optical tomogram. The  optical tomogram contains complete information about the system and can be considered as a primary object which describes the system completely, other than the conventional state vector or the density matrix of the system \cite{Ibort}. After this experiment \cite{smithey}, many nonclassical states of light have been characterised using optical homodyne tomography. A review of continuous-variable optical quantum-state tomography is given in \cite{lvovsky}.

Entangled states of light  have been recently\cite{Dauria,yao,lvovsky2,morin} characterised using optical homodyne tomography. In these experiments, the two-mode density matrix of the system has been reconstructed from the optical tomogram and the entanglement  is quantified using the reconstructed density matrix. A conditional measurement on one of the modes of entangled states may change the state in the other mode due to entanglement, and such changes may shows up in the optical tomogram of the state. In this paper, we study the optical tomography of bipartite continuous-variable entangled states of light. Main goal of this work is to find the signatures of entanglement in the optical tomogram of the state, without reconstructing the density matrix of the state. For this purpose, we investigate the optical tomogram of maximally entangled coherent states created in a beam splitter. The analytical results presented in this paper will be useful for the characterisation of the continuous-variable entangled states.

The optical tomogram of a quantum state can be theoretically evaluated using the symplectic tomograhic approach \cite{mancini1995,ariano1996,mancini1996,OVmanko1997,manko2005,manko2006}. The optical tomogram calculated by this method can be used to compare and verify the optical tomogram obtained by the homodyne detection of photonic states. The symplectic tomography of several single-mode nonclassical states of light have been investigated in the literature \cite{bazrafkan,filippov,korennoy,adam}. A theoretical investigation of the optical tomogram for two-mode coherent states of charged particle moving in a varying magnetic field is available in \cite{manko2012}. The correctness of measured tomogram can be checked using the properties of tomogram, like uncertainty relations \cite{nicola}, tomographic entropic inequalities \cite{manko2009}, purity constraints \cite{manko2011}, etc. An experimental check of the two-mode Robertson uncertainty relations and inequalities for the highest quadrature moments using homodyne photon detection have also been suggested \cite{manko2012b}. The operational use of experimentally measured optical tomograms to determine state characteristics avoiding the reconstruction of quasiprobabilities has been demonstrated in \cite{bellini2012}. We use the symplectic tomographic approach to find the optical tomogram of the two-mode photonic state. This paper is organized as follows. In section \ref{beamsplitting}, we discuss the generation of maximally entangled states using a beam splitter. Section \ref{OpticalTomography} is devoted to the discussion of optical tomography of entangled states generated at the output of the beam splitter. It also explains the signatures of the entanglement in the optical tomogram of the state. Section \ref{mandel} gives a verification for the different features shown by the optical tomogram by investigating the photon number statistics of the corresponding state. Finally in section \ref{conclusion}, we summarize the main results of this
paper.

\section{Beam splitting action}
\label{beamsplitting}
\begin{figure}
	\begin{center}
	\includegraphics[height=5 cm,width=5 cm]{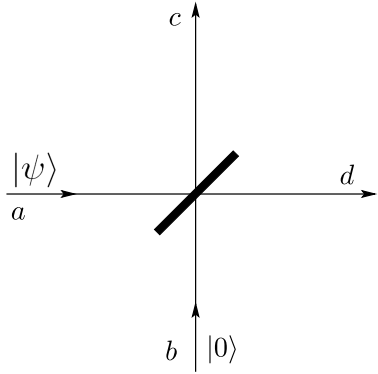} 
   \end{center}
   \caption{ \label{beamsplitter} A $50/50$ beam splitter with $\ket{\psi}$ in the horizontal port and $\ket{0}$ in the vertical port. $a$ and $b$ ($c$ and $d$) are the input (output) field modes. } 
   \end{figure}

Consider a $50/50$ beam splitter with zero phase difference between reflected and transmitted port. The unitary operator for the beam splitter reads 
\bea
U=\exp\left[\frac{\pi}{4}(a^\dag b-a b^\dag)\right],\label{BSoperator}
\eea
where $a$ and $b$ are the bosonic operators for the input field modes. The output field modes of the beam splitter are designated by $c$ and $d$. A schematic diagram of the beam splitter is given in fig.~\ref{beamsplitter}. A beam splitter generates entangled state if one of the input fields is nonclassical \cite{kim}. 
We consider both classical and nonclassical states  in the horizontal input port (mode $a$) of the beam splitter and study the optical tomogram of the output states. In both of these cases we take vacuum state $\ket{0}$ in the vertical input port (mode $b$) of the beam splitter. The states considered for classical and nonclassical states are  coherent state, and  even and odd coherent states, respectively.  The initial even and odd coherent states exhibits different nonclassical behaviour during the time evolution when compared to an initial coherent state \cite{rohith}. Beam splitting action on the coherent state $\ket{\psi}=\ket{\alpha}$, where $\alpha$ ($=\modu{\alpha}e^{i\delta}$) is a complex number, with vacuum $\ket{0}$ generates the separable state 
\bea
\ket{\Phi}_{cs}=\ket{\beta}\otimes\ket{\beta},
\eea
where $\beta=\alpha/\sqrt{2}$. Next, we consider beam splitting of even and odd coherent states, defined by 
\bea
\ket{\psi}_{h}=N_{h}\left[ \ket{\alpha}+e^{i\pi h}\ket{-\alpha}\right], \label{initial}
\eea
where the normalisation constant
\bea
N_{h}=\frac{1}{\sqrt{2}}\left[1+(-1)^h\, e^{-2\modu{\alpha}^2}\right]^{-\frac{1}{2}}.
\eea
Here $h=0$ and $1$ corresponds to the even and odd coherent states, respectively. In this case, we get entangled states at the output modes of the beam splitter. The state $\ket{\Phi}_{h}$ of the beam splitter output modes is calculated using the unitary operator given in eq.~(\ref{BSoperator}):
\bea
\ket{\Phi}_{h}=N_{h}\left[\ket{\beta}_c\ket{\beta }_d+e^{i\pi h}\ket{-\beta}_c\ket{-\beta }_d\right].\label{entstate}
\eea
In the limit of large $\modu{\alpha}^2$, the coherent states appearing in the superposition given in eq.~(\ref{initial}) form an orthogonal basis and thus the entangled state given in eq.~(\ref{entstate}) can be written in the Schmidt decomposition form \cite{vanEnk}. Since all the Schmidt coefficients have the same magnitude $1/\sqrt{2}$, the entanglement of the state $\ket{\Phi}_{h}$ can be found to be $E=\log_2 (2)=1$ ebits, which is the maximum entanglement possible in $2$ dimension. Hence, for large $\modu{\alpha}^2$, the state $\ket{\Phi}_{h}$ is a maximally entangled state in $2$ dimension.

\section{Optical tomography}
\label{OpticalTomography}
We theoretically calculate   the optical tomogram of the  maximally entangled state $\ket{\Phi}_{h}$  and  separable state $\ket{\Phi}_{cs}$ using the symplectic tomogram of the state. Consider the homodyne quadratures $\hat{X}_{\mu_1,\nu_1}$ and $\hat{X}_{\mu_2,\nu_2}$ associated with mode $c$ and mode $d$ respectively:
\begin{eqnarray}
\begin{aligned}[c]
\hat{X}_{\mu_1,\nu_1}= \mu_1\, \hat{q}_1+\nu_1\, \hat{p}_1,  
\end{aligned}
\quad
\begin{aligned}[c]
\hat{X}_{\mu_2,\nu_2}=\mu_2\, \hat{q}_2+\nu_2\, \hat{p}_2.
\end{aligned}
\end{eqnarray}
Here $\mu_1$, $\nu_1$, $\mu_2$, $\nu_2$ are real parameters and $\hat{q}$s and $\hat{p}$s are the usual position and momentum operators. The symplectic tomogram of a two-mode state with density matrix $\rho$ is defined as \cite{ariano1996,Ibort}
\begin{eqnarray}
\begin{split}
M&(X_{\mu_1,\nu_1},\mu_1, \nu_1;X_{\mu_2,\nu_2},\mu_2,\nu_2)\\&=\text{Tr}\left[\rho\, \delta(X_{\mu_1,\nu_1}\hat{I}-\hat{X}_{\mu_1,\nu_1})\,\delta(X_{\mu_2,\nu_2}\hat{I}-\hat{X}_{\mu_2,\nu_2})\right], 
\end{split}
\end{eqnarray}
where $X_{\mu_1,\nu_1}$ and $X_{\mu_2,\nu_2}$ are the eigenvalue of the Hermitian operators $\hat{X}_{\mu_1,\nu_1}$ and $\hat{X}_{\mu_2,\nu_2}$, respectively. For a pure two-mode state with wave vector $\ket{\Phi}$, the symplectic tomogram can be obtained by taking the fractional Fourier transform of its coordinate wave function $\Phi(x,y)$ \cite{manko1999,manko2012}:
\begin{eqnarray}
\begin{split}
M&( X_{\mu_1,\nu_1},\mu_1,\nu_1; X_{\mu_2,\nu_2},\mu_2,\nu_2)=\frac{1}{4\pi^2\modu{\nu_1 \nu_2}}\times\\
&\left|\int dx\,dy\, \Phi(x,y) \exp\left(\frac{i \mu_1}{2\nu_1}x^2-\frac{i{X_{\mu_1,\nu_1}}}{\nu_1}x\right)\right.\\
&\left.\times\exp\left(\frac{i \mu_2}{2\nu_2}y^2-\frac{i{X_{\mu_2,\nu_2}}}{\nu_2}y\right)\right|^2,
\end{split} \label{sym_tomo}
\end{eqnarray}
In experiments optical tomogram of the states are generated and we look for the signatures of entanglement in the optical tomogram.
 The optical tomogram of a two-mode state can be obtained from the symplectic tomogram given in eq.~(\ref{sym_tomo}) by the following substitutions \cite{mancini1995,ariano1996,manko1999,manko2012}:
\begin{eqnarray}
X_{\mu_1,\nu_1}=X_{\theta_1},\q \mu_1=\cos \theta_1,\q \nu_1=\sin \theta_1,\nonumber\\
X_{\mu_2,\nu_2}=X_{\theta_2},\q \mu_2=\cos \theta_2,\q \nu_2=\sin \theta_2.
\end{eqnarray}
Here $X_{\theta_1}$ and $\theta_1$ ($X_{\theta_2}$  and $\theta_2$) are the quadrature and the phase of local oscillator in homodyne detection setup for mode $c$ (mode $d$), respectively. The phase of the local oscillators varies in the domain $0\leq \theta_1,\theta_2 \leq 2\pi$. The optical tomogram of the two-mode state is then reads as
\begin{eqnarray}
\begin{split}
\omega & \left(X_{\theta_1},\theta _1; X_{\theta_2},\theta _2\right)\\
&=M\left(X_{\theta_1},\cos\theta_1,\sin\theta_1;X_{\theta_2},\cos\theta_2,\sin\theta_2\right).\label{transform}
\end{split}
\end{eqnarray} 
Two-mode coordinate wave function $\Phi_{h}(x,y)$ for the entangled state $\ket{\Phi}_{h}$ is calculated as
\begin{eqnarray}
\begin{split}
\Phi_{h}(x,y)&=\frac{N_{h}}{\sqrt{\pi}}\sum_{r=0}^{1} e^{-i\pi r h} \times\\ &\exp(-\modu{\beta}^2/2-{\beta^2_r}/2+\sqrt{2}\beta_r x-x^2/2) \times\\
&\exp(-\modu{\beta}^2/2-{\beta^2_r}/2+\sqrt{2}\beta_r y-y^2/2),
\end{split}
\end{eqnarray} 
where $\beta_r=\beta e^{i\pi r}$. Substituting $\Phi_{h}(x,y)$ in eq.~(\ref{sym_tomo}) and using the transformation given in eq.~(\ref{transform}), the two-mode optical tomogram for the entangled state $\ket{\Phi}_{h}$ is obtained as
\bea
\begin{split}
&\omega_{h}(X_{\theta_1},\theta _1;X_{\theta_2},\theta _2)\\&=\frac{N^2_{h}}{\pi}\modu{\sum_{r=0}^{1}e^{-i\pi r h} ~\eta_r(X_{\theta_1},\theta_1)\eta_r(X_{\theta_2},\theta_2)}^2, \label{tomogram}
\end{split}
\eea
where
\bea
\begin{split}
\eta_r(X_{\theta_i},\theta_i)=\exp&\left(-\frac{\modu{\beta}^2}{2}-\frac{{X_{\theta_i}}^2}{2}\right.\\
&\left.+\sqrt{2}\beta_r X_{\theta_i} e^{-i{\theta_i}}- \frac{\beta^2_r e^{-i{2\theta_i}}}{2}\right),
\end{split}
\eea
for $i=1$ and $2$.

Now, for the separable two-mode state $\ket{\Phi}_{cs}=\ket{\beta}_c\ket{\beta}_d$ the optical tomogram can be written as the product of optical tomograms of the subsystems \cite{Ibort}, that is $\omega_{cs}(X_{\theta_1},\theta _1;X_{\theta_2},\theta _2)=\omega_1(X_{\theta_1},\theta _1)\times\omega_2(X_{\theta_2},\theta _2)$, where $\omega_1(X_{\theta_1},\theta _1)$, $\omega_2(X_{\theta_2},\theta _2)$ are the optical tomogram for mode $c$ and mode $d$, respectively. It can be shown that
\bea
\omega_i(X_{\theta_i},{\theta_i})=\frac{1}{\sqrt{\pi}}\exp\left[-\left(X_{\theta_i}-\sqrt{2}\modu{\beta}\cos(\delta-{\theta_i})\right)^2\right],\label{subsystemtomo}
\eea
where $i =1$ or $2$. The maximum intensity of the optical tomogram $\omega_{i}(X_{\theta_i},{\theta_i})$ is $1/\sqrt{\pi}$, which occurs along a sinusoidal path, defined by $X_{\theta_i}=\sqrt{2 \modu{\beta}^2} \cos(\delta-{\theta_i})$, in the $X_{\theta_i}-{\theta_i}$ plane. Hence the optical tomogram in mode $c$ (mode $d$) will always be a sinusoidal single-stranded structure for any values of parameters in mode $d$  (mode $c$) and $\delta$.
 
\subsection{Signature of entanglement}
The state $\ket{\Phi}_h$ is an entangled state and the corresponding optical tomogram cannot be written in the form of the product of subsystem tomograms, that is $\omega_{h}(X_{\theta_1},\theta _1;X_{\theta_2},\theta _2)\neq \omega_1(X_{\theta_1},\theta _1)\times\omega_2(X_{\theta_2},\theta _2)$. We present our analysis for $h=0$, that is, for initial even coherent state. When $h=0$, the entangled output state $\ket{\Phi}_{0}=N_{0}\left[ \ket{\beta }\ket{\beta}+\ket{-\beta }\ket{-\beta}\right]$. A measurement of field quadrature in any of the modes will collapse the entanglement between the modes. That is, the quantum state in one mode is correlated with the quadrature measurement in the other mode. For example, a measurement of quadrature $\hat{X}_{\theta_2}$ in mode $d$ will project the state $\ket{\Phi}_{0}$ to the state $\ket{\phi}_{0,c}$ in mode $c$:
\bea
\ket{\phi}_{0,c}=\tilde{N}_{0}\left[\psi_{\beta}(X_{\theta_2},\theta_2)\ket{\beta}+\psi_{-\beta}(X_{\theta_2},\theta_2)\ket{-\beta}\right],\label{phi_c}
\eea
where $\tilde{N}_{0}$ is appropriate normalisation constant and $\psi_{\pm \beta}(X_{\theta_2},\theta_2)=\langle X_{\theta_2},\theta_2 \ket{\pm \beta}$, is the quadrature representation of the coherent state $\ket{\pm \beta}$. Based on the relative strength of the coefficients $\psi_{\beta}(X_{\theta_2},\theta_2)$ and $\psi_{-\beta}(X_{\theta_2},\theta_2)$,  the state $\ket{\phi}_{0,c}$ can be one of following: $\ket{\beta}$,  $\ket{-\beta}$ and  a superposition of $\ket{\beta}$ and  $\ket{-\beta}$. For large $\modu{\alpha}^2$, the probability for occurring the state $\ket{\pm \beta}$ is proportional to 
\bea
\begin{split}
\modu{\psi_{\pm \beta}(X_{\theta_2},\theta_2)}^2 &=\frac{1}{\sqrt{\pi}}\exp\left[-\modu{\beta}^2-2\modu{\beta}^2\cos 2(\delta-\theta_2)\right.\\
&\left.-X^2_{\theta_2} \pm 2\sqrt{2}X_{\theta_2}\modu{\beta}\cos (\delta-\theta_2) \right].
\end{split}
\label{psi_beta}
\eea

\begin{figure*}
   \begin{center}
	\includegraphics[height=5 cm,width=5 cm]{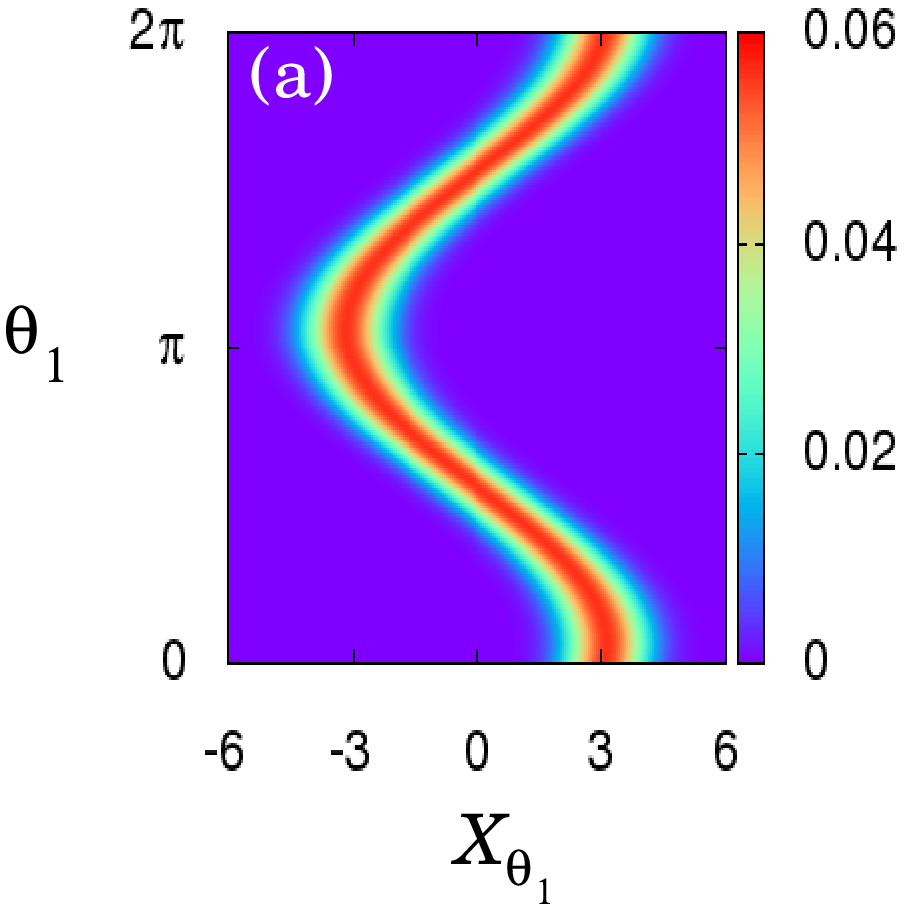} \quad \includegraphics[height=5 cm,width=5 cm]{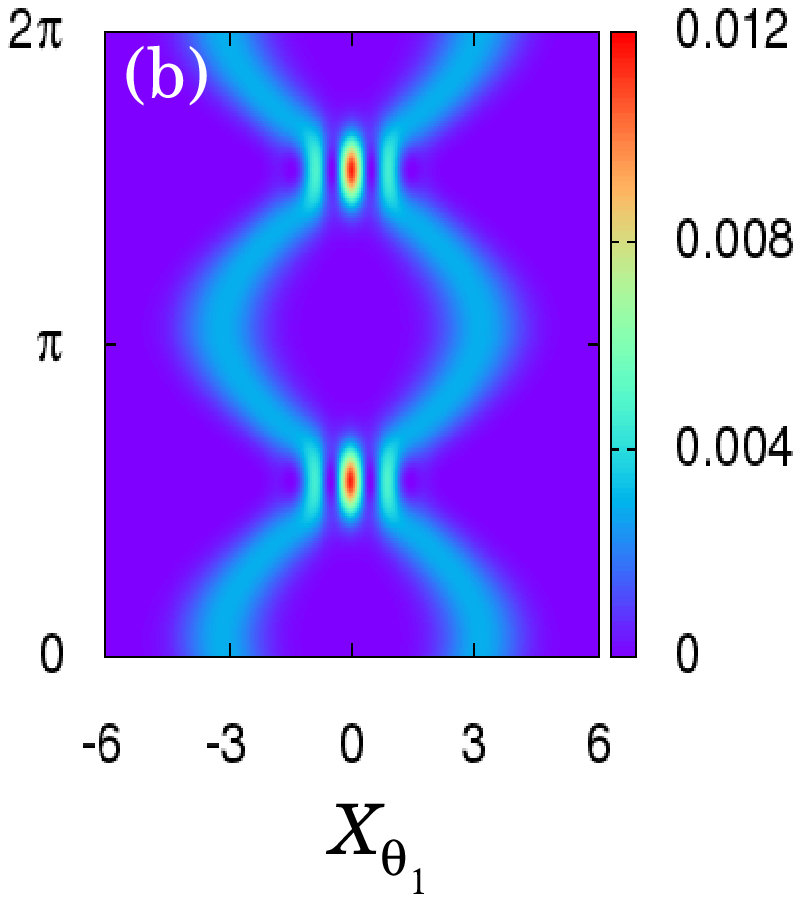}\quad \includegraphics[height=5 cm,width=5 cm]{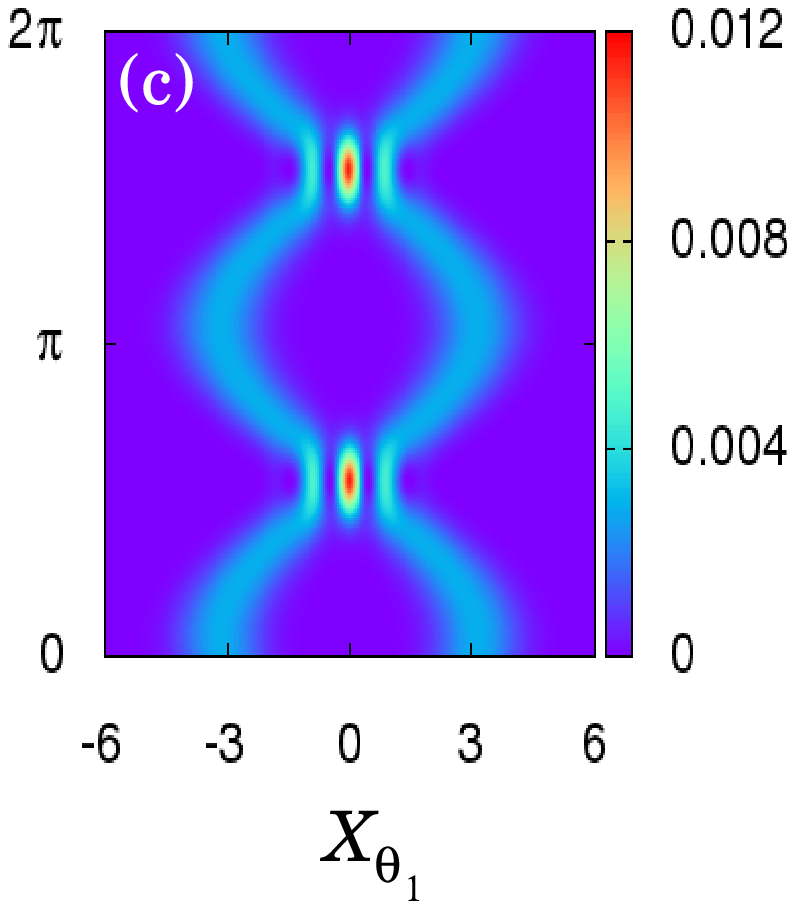}
   \end{center}
   \caption 
   { \label{fig:l2tomo} Optical tomograms $\omega_{0}(X_{\theta_1},\theta _1;X_{\theta_2},\theta _2)$ in mode $c$ for the entangled state $\ket{\Phi}_0$ with $\modu{\alpha}^2=10$, $\delta=0.2$, and $X_{\theta_2}=2.0$, for different relative phases $\modu{\delta-\theta_2}$ of the quadrature measurement in mode $d$: (a) $0.3$ (b) $\pi/2$ and (c) $3\pi/2$, respectively. Optical tomogram is a sinusoidal single-stranded structure for $\modu{\delta-\theta_2}=0.3$. For $\modu{\delta-\theta_2}=\pi/2$ and $3\pi/2$, the optical tomogram shows sinusoidal double-stranded structure.} 
   \end{figure*}

For $X_{\theta_2}\neq 0$, relative strength of the probabilities crucially depend on the last term in eq.~(\ref{psi_beta}). In the range $\pi/2< \modu{\delta-\theta_2} < 3\pi/2$,  the state $\ket{\phi}_{0,c}$ becomes a coherent state $\ket{-\beta}$ because $\modu{\psi_{-\beta}(X_{\theta_2},\theta_2)}^2 \gg \modu{\psi_{\beta}(X_{\theta_2},\theta_2)}^2$.  Thus, the optical tomogram in mode $c$ will be a single-stranded structure corresponds to the coherent state $\ket{-\beta}$ (Note that the optical tomogram for a coherent state $\ket{\beta}$ in $X_{\theta_i}-\theta_i$ plane is a sinusoidal single-stranded structure as given in eq.~(\ref{subsystemtomo})). Also, in the range $0\leq \modu{\delta-\theta_2}< \pi/2$ and $3\pi/2< \modu{\delta-\theta_2}\leq 2\pi$, the state $\ket{\phi}_{0,c}$ becomes a coherent state $\ket{\beta}$ because  $\modu{\psi_{\beta}(X_{\theta_2},\theta_2)}^2 \gg \modu{\psi_{-\beta}(X_{\theta_2},\theta_2)}^2$, which gives   a single-stranded structure for the optical tomogram in mode $c$. Figure \ref{fig:l2tomo}(a) shows single-stranded structure in the optical tomogram $\omega_{0}(X_{\theta_1},\theta _1;X_{\theta_2},\theta _2)$ 
for  $\modu{\alpha}^2=10$, $\delta=0.2$,  $X_{\theta_2}=2.0$ and $\modu{\delta-\theta_2}=0.3$.

 It can be shown that $\modu{\psi_\beta(X_{\theta_2},\theta_2)}^2=\modu{\psi_{-\beta}(X_{\theta_2},\theta_2)}^2$ for $\modu{\delta-\theta_2}=\pi/2$ or $3\pi/2$ (within the periodicity of optical tomogram). For these two values of $\modu{\delta-\theta_2}$, the probability for occurring $\ket{\beta}$ and $\ket{-\beta}$ in mode $c$ is $50:50$ and hence the state $\ket{\phi}_c$ reduces to even coherent state of the form $\left[\ket{\beta}+\ket{-\beta}\right]$. The optical tomogram in mode $c$ will display double-stranded structure, in which, one strand corresponds to $\ket{\beta}$ and the other corresponds to $\ket{-\beta}$. The optical tomogram in mode $c$ for $\modu{\delta-\theta_2}=\pi/2$ and $3\pi/2$ with $X_{\theta_2}=2.0$ are shown in figs.~\ref{fig:l2tomo}(b) and \ref{fig:l2tomo}(c), respectively.
 Quantum interference between the state $\ket{\beta}$ and $\ket{-\beta}$ are reflected in the optical tomogram at regions in $X_{\theta_1}-\theta_1$ plane, where the two strands intersect.  When $X_{\theta_2}=0$, the state $\ket{\phi}_{0,c}$ is always an even coherent state, without any condition on  $\modu{\delta-\theta_2}$. This displays double-stranded structure in the optical tomogram. 

\section{Mandel's $Q$ parameter}
\label{mandel}
\begin{figure}
      \begin{center}
	\includegraphics[height=5 cm,width=6 cm]{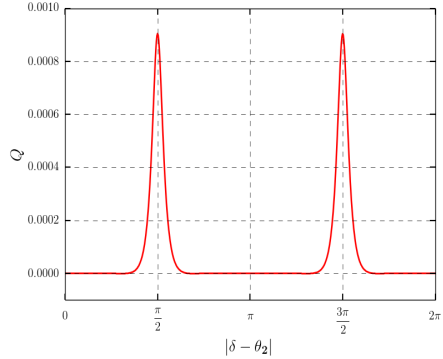} 
   \end{center}
  
   \caption{  \label{fig:mandel}Mandel's $Q$ parameter for the state $\ket{\phi}_{0,c}$ as a function of the relative phase $\modu{\delta-\theta_2}$ of the measurement in mode $d$ with $\modu{\alpha}^2=10$. The positive value of $Q$ at $\modu{\delta-\theta_2}=\pi/2$ and $3\pi/2$ (and close to it) indicates the super-Poissonian statistics of the state $\ket{\phi}_{0,c}$ and for all other values of $\modu{\delta-\theta_2}$, the state $\ket{\phi}_{0,c}$ exhibit Poissonian statistics ($Q=0$).} 
   \end{figure}

\begin{figure*}
   \begin{center}
	\includegraphics[height=5 cm,width=5 cm]{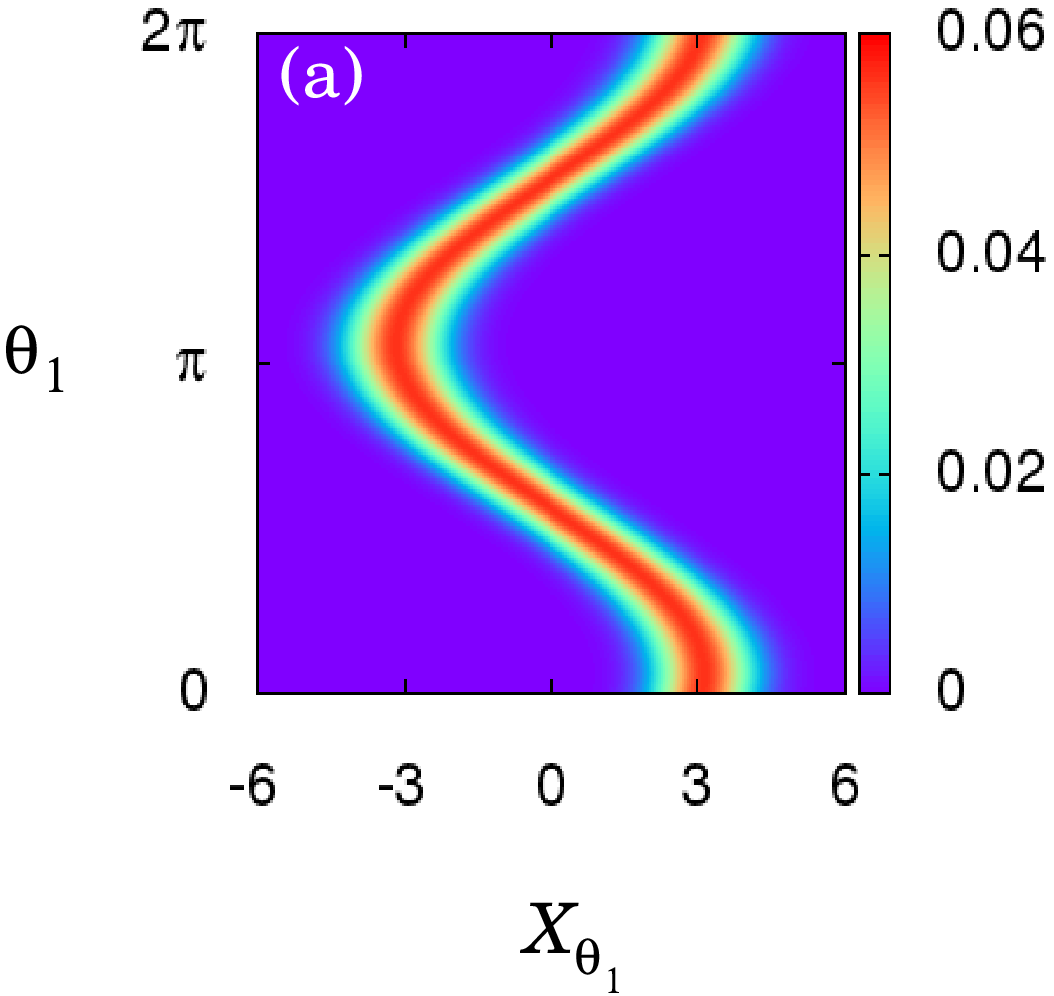} \quad \includegraphics[height=5 cm,width=5 cm]{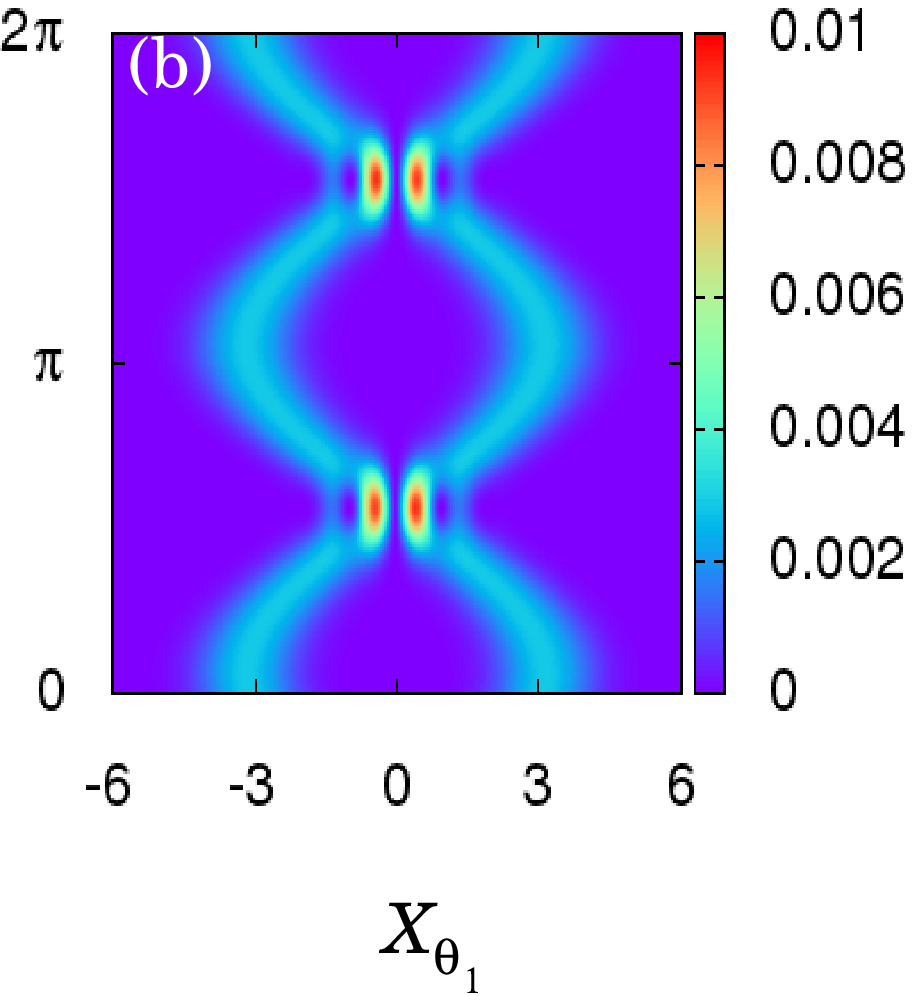}\quad \includegraphics[height=5 cm,width=5 cm]{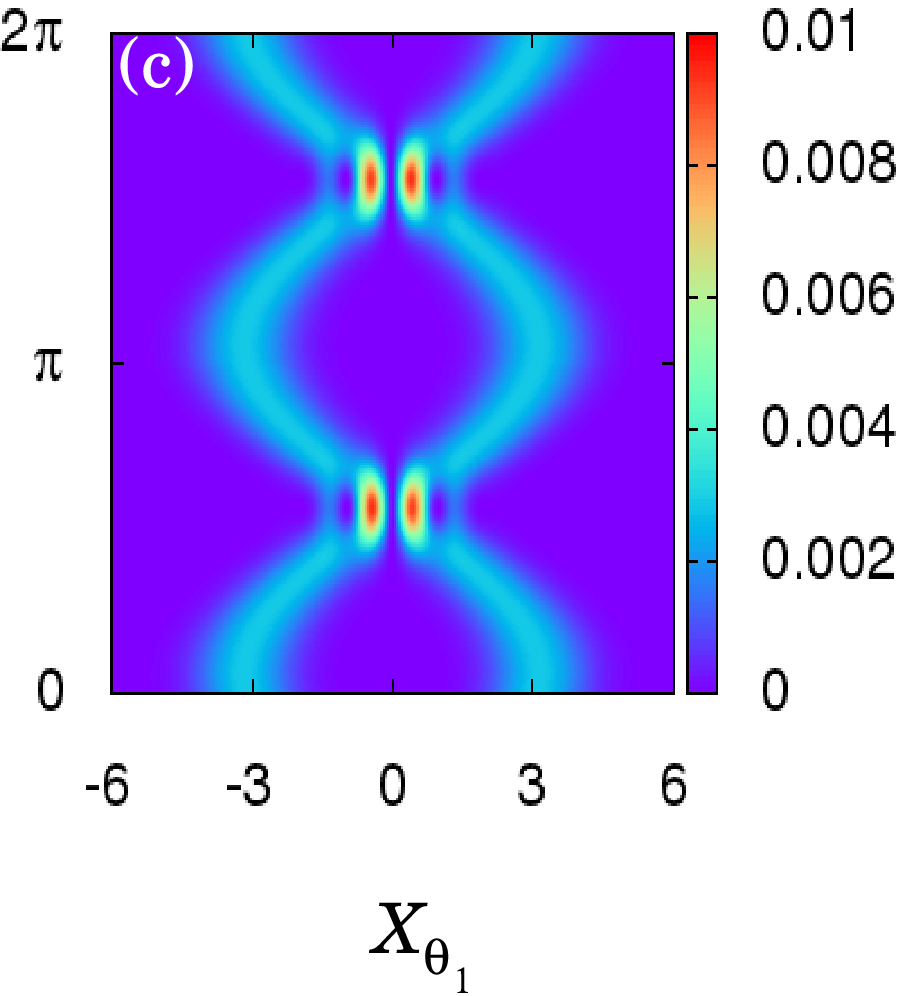}
   \end{center}
   \caption 
   { \label{fig:l2h1tomo} Optical tomograms $\omega_{1}(X_{\theta_1},\theta _1;X_{\theta_2},\theta _2)$ in mode $c$ for the entangled state $\ket{\Phi}_{1}$ with $\modu{\alpha}^2=10$, $\delta=0.2$, and $X_{\theta_2}=2.0$, for different relative phases $\modu{\delta-\theta_2}$ of the quadrature measurement in mode $d$: (a) $0.3$ (b) $\pi/2$ and (c) $3\pi/2$, respectively. Optical tomogram is a sinusoidal single-stranded structure for $\modu{\delta-\theta_2}=0.3$. For $\modu{\delta-\theta_2}=\pi/2$ and $3\pi/2$, the optical tomogram shows sinusoidal double-stranded structure.} 
   \end{figure*}

We can also show the above features using the statistics of photon number in the state $\ket{\phi}_{0,c}$, specifically, in terms of the Mandel's $Q$ parameter, defined as \cite{mandel}
\bea
Q=\frac{\aver{\hat{n}^2}-\aver{\hat{n}}^2}{\aver{\hat{n}}}-1,
\eea
where $\hat{n}$ is the photon number operator. A positive value of $Q$ indicates the super-Poissonian statistics of the field, and $Q=0$ indicates the Poissonian statistics exhibited by a coherent field. $Q$ parameter of the state $\ket{\phi}_{0,c}$ in mode $c$ as a function of relative phase $\modu{\delta-\theta_2}$, is plotted in fig.~\ref{fig:mandel}. It shows that at $\modu{\delta-\theta_2}=\pi/2$, $3\pi/2$ and close to it, the state exhibits super-Poissonian statistics and for all other $\modu{\delta-\theta_2}$ values, the state $\ket{\phi}_{0,c}$ shows Poissonian statistics corresponding to a coherent field, which is either $\ket{\beta}$ or $\ket{-\beta}$. The different statistics of photon number exhibited by the state in mode $c$ upon changing the parameters in mode $d$, has been experimentally observed in the case of micro-macro entanglement of light \cite{lvovsky2} using the reconstructed density matrix. We have theoretically shown that, without reconstructing the density matrix of the system, the signature of entanglement can be directly observed in  the optical tomogram of the state. When the initial state is an  odd coherent state (i.e., $h=1$), we get the entangles state $\ket{\Phi}_{1}$ given in eq. ~(\ref{entstate}) at the  output modes of the beam splitter.  We have repeated the forgoing analysis for the entangled state $\ket{\Phi}_{1}$ and verified the results obtained earlier. The    figs.~\ref{fig:l2h1tomo} (a)-(c) shows the optical tomograms   $\omega_{1}(X_{\theta_1},\theta _1;X_{\theta_2},\theta _2)$  in mode $c$ for the entangled state $\ket{\Phi}_{1}$ with same set of parameters used in the case of the entangled state $\ket{\Phi}_{0}$.

\section{Conclusion}
\label{conclusion}
A closed form  analytical expression for the optical tomogram of the maximally entangled coherent state generated at the output of the beam splitter is derived by taking the fractional Fourier transform of the output two-mode wave function. For separable two-mode states, the optical tomogram of the system can be written as the product of the optical tomograms of the subsystems. Whereas, for the entangled two-mode states, the optical tomogram in the mode $c$ shows different features when we change  the parameters  $X_{\theta_2}$ and $\theta_2$ in the mode $d$. Similarly, the optical tomogram in the mode $d$ will be affected by the parameters in mode $c$. Specifically, for the entangled state $\ket{\Phi}_{h}=N_{h}\left[\ket{\beta}\ket{\beta}+e^{i\pi h} \ket{-\beta}\ket{-\beta}\right]$, with $X_{\theta_2}\neq 0$, the optical tomogram in mode $c$ shows double-stranded structure if $\modu{\delta-\theta_2}=\pi/2$ or $3\pi/2$ and a single-stranded structure for all other values of $\modu{\delta-\theta_2}$. Our analytical result for the optical tomogram of entangled coherent states may be useful for the experimental characterization of such states.

\end{document}